\documentclass[aps,prd,preprintnumbers,twocolumn,superscriptaddress,showpacs,letter,nofootinbib]{revtex4-1}
\usepackage{graphicx}
\usepackage{dcolumn}
\usepackage{bm}
\usepackage{amsmath}
\usepackage{units}
\usepackage{lineno}
\usepackage{cancel}
\usepackage{color}

\usepackage{amssymb}
\newcommand{\ba}{\begin{array}}
\newcommand{\ea}{\end{array}}

\def\br{\begin{eqnarray}}
\def\er{\end{eqnarray}}
\def\be{\begin{equation}}
\def\ee{\end{equation}}

\def\({\left(}
\def\){\right)}

\def\lapproxeq{\lower .7ex\hbox{$\;\stackrel{\textstyle                                                    
<}{\sim}\;$}}                                                    
\def\gapproxeq{\lower .7ex\hbox{$\;\stackrel{\textstyle                                                    
>}{\sim}\;$}}                                                    
\def\be{\begin{equation}}                                                    
\def\ee{\end{equation}}                                                    
\def\bea{\begin{eqnarray}}                                                    
\def\eea{\end{eqnarray}}

\def\sh{\hat s}
\def\sh2{{\hat s}^2}

\begin{document}


\title{Multiplicity distributions in the eikonal and the\\ $U$-matrix unitarization schemes}

\author{E.~G.~S.~Luna}
\email{luna@if.ufrgs.br}
\affiliation{Instituto de F\'isica, Universidade Federal do Rio Grande do Sul, Caixa Postal 15051, 91501-970, Porto Alegre, RS, Brazil}
\author{M.~G.~Ryskin}
\email{ryskin@thd.pnpi.spb.ru}
\affiliation{Petersburg Nuclear Physics Institute, NRC Kurchatov Institute, \\ Gatchina, St.~Petersburg, 188300, Russia}


\begin{abstract}

  The multiplicity distribution of charged particles produced in the central rapidity ($|\eta|<2.5$) region is calculated for the eikonal and the $U$-matrix unitarization schemes using the AGK cutting rules and compared with the ATLAS 13 TeV data. The data favours the eikonal approach.
  
\end{abstract}


\maketitle

\section{Introduction}

Studying the high energy behaviour of the $ pp$ amplitude is one of the goals of the LHC. Since the total cross section grows with the energy it is natural to take the QCD Pomeron (or the soft Pomeron with the intercept $\alpha_{\Bbb P}(0)>1$) as the seed Born amplitude. 
However, at asymptotically high energy, $\sqrt s\to\infty$, the power growth of this Born amplitude violates the Froissart limit $\sigma<C\cdot\ln^2s$~\cite{Fr}.
 To satisfy unitarity the amplitude, $A(s)$, should be unitarized. 
 
There are two possibilities of the unitarization based on two solutions of the $s$-channel two-particle unitarity equation
\begin{equation}
\label{un}
2\mbox{Im}{\cal A}(s,b) = |{\cal A}(s,b)|^2+G_{inel}(s,b),
\end{equation}
written in impact parameter, $b$, representation. Here ${\cal A}$ is the elastic scattering  amplitude and  a real non-negative quantity, $G_{inel}$, describes the inelastic channels contribution. 

Denoting the ratio of real to imaginary part of ${\cal A}$ by $\rho(s,b)=\mbox{Re}/\mbox{Im}$ we get the solutions of (\ref{un}) in the form
\begin{equation}
\mbox{Im}{\cal A}(s,b)=\frac{1\pm\sqrt{1-(1+\rho^2)G_{inel}(s,b)}}{1+\rho^2},
\label{sol}
\end{equation}
where
\begin{equation}
0 \leq G_{inel}(s,b) \leq (1 + \rho^{2})^{-1} ,
\label{unitarity004}
\end{equation}
since $\mbox{Im}{\cal A}(s,b)$ is real. 

The eikonal unitarization corresponds to the solution of equation (\ref{sol}) with the minus sign, i.e., the one with
the negative square root. In this case, the $pp$-amplitude reads as
\begin{equation}
{\cal A}(s,t) = s\int bdbJ_0(bq){\cal A}(s,b)
\end{equation}
with 
\begin{widetext}
\begin{equation}
  \label{eik}
        {\cal A}(s,b)~=~i[1-e^{i\chi(s,b)}] = -i\sum_{n=1}^\infty \frac{[i\chi(s,b)]^{n}}{n!} =
        i\sum_{n=1}^\infty C_{n}^{eik} \cdot (-1)^{n-1} [\Omega(s,b)]^{n},  
\end{equation}
\end{widetext}
where the function $\chi(s,b)$ corresponds to the one-Pomeron exchange, $C_{n}^{eik} = 2^{-n}/n!$, 
$\Omega(s,b)\equiv -2i\chi(s,b)$ is the opacity of $pp$ interaction, $J_0(x)$ is the Bessel function of the first kind, and $q=\sqrt{-t}$ is the momentum transferred. \\

In the eikonal scheme, solving the unitarity equation (\ref{un}) for $G_{inel}(s,b)$ in terms of the function $\chi(s,b)$ yields
\begin{equation}
\label{inel}
G_{inel}(s,b) = 1-e^{-2 \textnormal{Im}\chi(s,b)} = 1-e^{- \textnormal{Re}\Omega(s,b)} .
\end{equation}

\begin{figure*}[t]
\begin{center}
\hspace{-0.4cm}
\includegraphics[trim=0.0cm 0cm 0cm 0cm,scale=0.69]{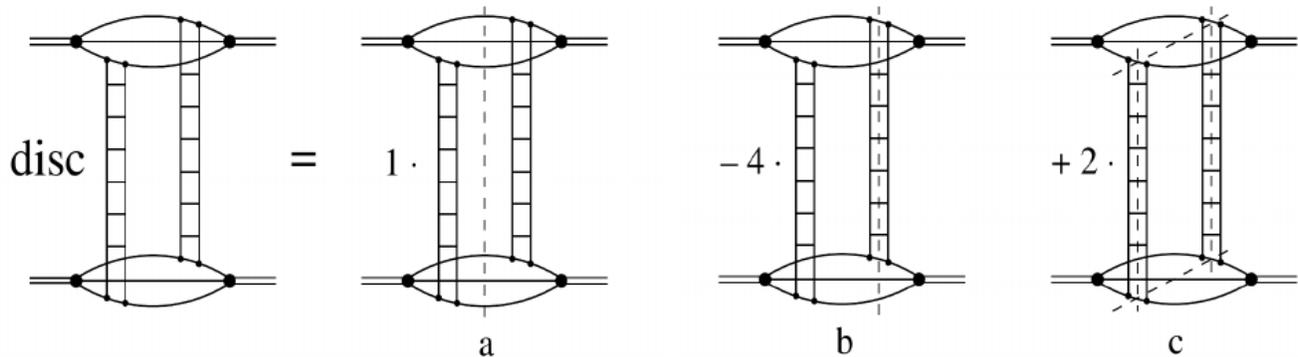}
\vspace{-0.5cm}
\caption{\sf Two-Pomeron exchange in the t channel expressed as a sum over all intermediate states in the s-channel.
}
\label{f1}
\end{center}
\end{figure*} 

The $U$-matrix unitarization (see e.g. \cite{TT,TT2}) corresponds to the solution
of  unitarity equation (\ref{un}) with the plus sign, i.e., the
one with the positive square root in (\ref{sol}). It leads  to the relation 

\begin{widetext}
\begin{equation}
  {\cal A}(s,b)~=~\frac{\hat{\chi}(s,b)}{1-i\hat{\chi}(s,b)/2} = -2i \sum_{n=1}^\infty \frac{[i\hat{\chi}(s,b)]^{n}}{2^{n}} =
  i\sum_{n=1}^\infty C_{n}^{U} \cdot (-1)^{n-1} [\hat{\Omega}(s,b)]^{n} ,
\label{u}  
\end{equation}
\end{widetext}
were the function $\hat{\chi}(s,b)$ corresponds to the one-Pomeron exchange in the $U$-matrix scheme, $C_{n}^{U} = 2/4^{n}$, and $\hat{\Omega}(s,b)\equiv -2i\hat{\chi}(s,b)$ is the respective opacity.
In the expansion over the $\Omega$ or $\hat{\Omega}$ powers in (\ref{eik},\ref{u}) each $\Omega^n$ or $\hat{\Omega}^n$ term corresponds to the exchange of $n$ Pomerons.

Note that the first two terms in the eikonal (\ref{eik}) 
and the $U$-matrix (\ref{u}) schemes (with $\Omega = {\Bbb P} = \hat{\Omega} $) are exactly the same.

The collider ($\sqrt s\geq 200$ GeV) data on total, $\sigma_{tot}$ and differential, $d\sigma_{el}/dt$ low $|t|<0.1$ GeV$^2$ cross sections and the Re/Im=$\rho(t=0)$ ratio were analysed in \cite{Cud,LMP} papers in terms of the eikonal and $U$-matrix unitarization schemes. It turns out that the available, $\sigma_{tot}, d\sigma_{el}/dt$ and $\rho$ data are not sufficient to distinguish between the $U$-matrix and the eikonal approaches. 

In the present paper, using the AGK Reggeon cutting rules~\cite{AGK}, we analyse the multiplicity distributions generated by the eikonal and the $U$-matrix multi-Pomeron processes.\footnote{The multiplicity distributions in the $U$-matrix and eikonal approaches were recently considered in \cite{OT} using a geometrical model. The authors assume a Negative Binomial Distribution for the elementary string and adopt an expression from \cite{luna5} for the mean multiplicity generated by a few strings. In contrast, in the present paper, we use explicit AGK rules in both the $U$-matrix and eikonal schemes to calculate the distribution over the number of cut Pomerons, which drives the final hadron multiplicities.} The AGK rules were proved in~\cite{AGK} and confirmed for the QCD case in~\cite{BR,BV}.

\begin{figure*}[t]
\begin{center}
\vspace{-3.5cm}
\includegraphics[scale=0.30]{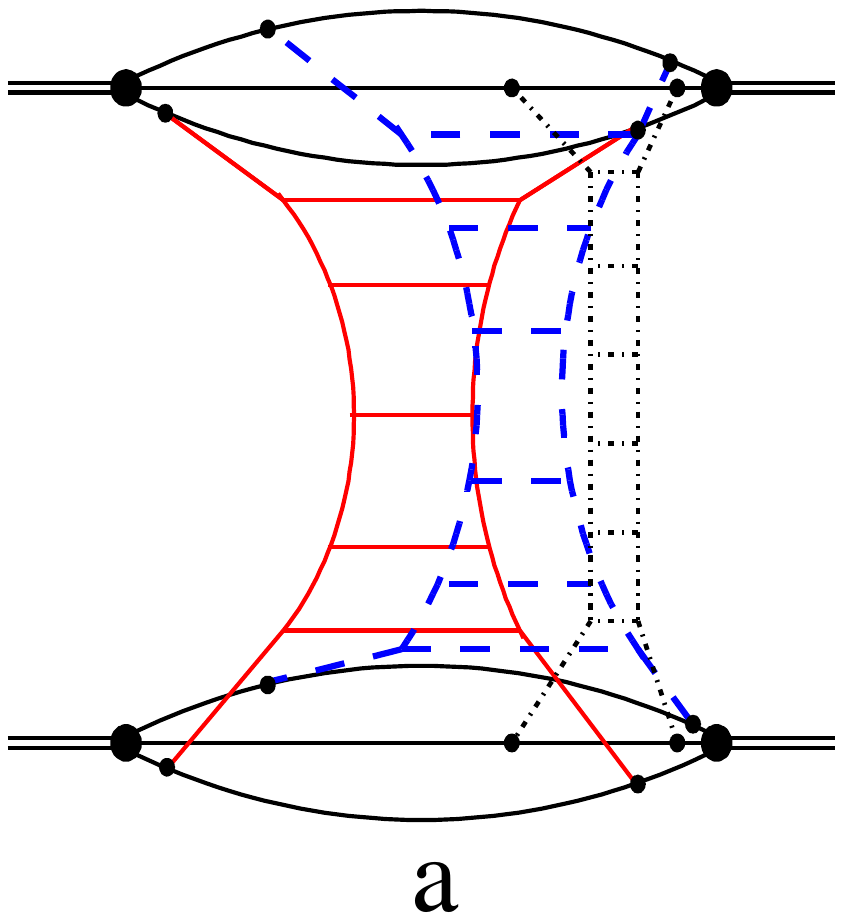}
\hspace{-0.5cm}
\includegraphics[scale=0.30]{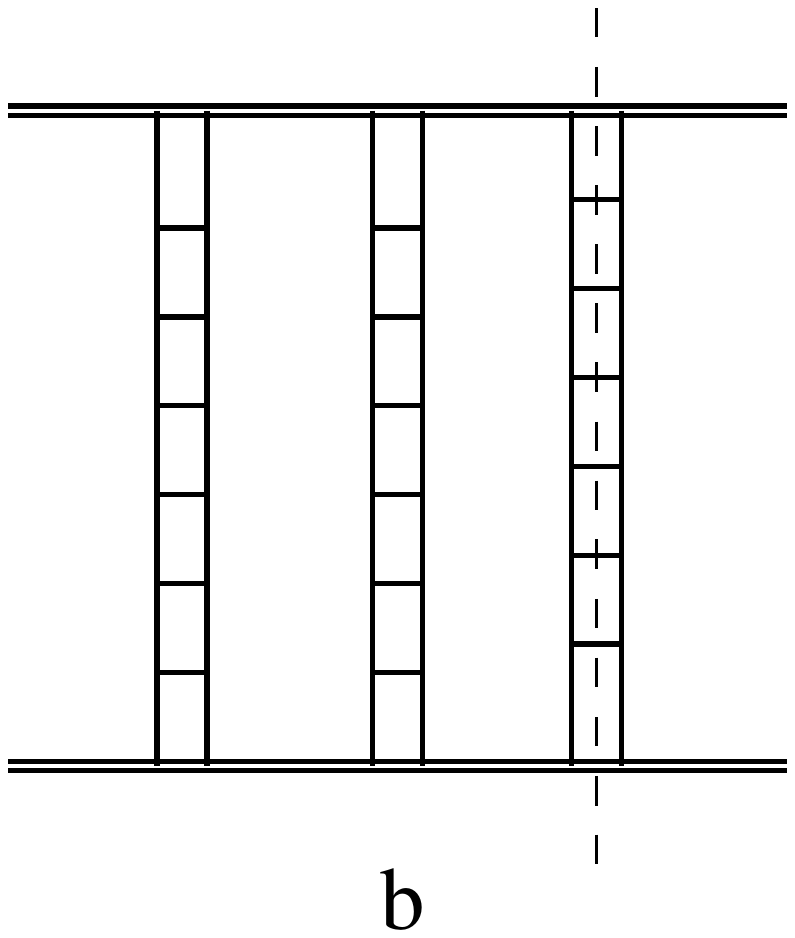}
\vspace{-0.5cm}
\caption{\sf Three-Pomeron exchange in the eikonal (a) and the $U$-matrix (b) unitarization schemes. The horizontal position of a particular point (line) reflects the time corresponding to this point.
}
\label{f2}
\end{center}
\end{figure*}

In Sec. 2 we recall the AGK rules and describe their structure for the eikonal and the $U$-matrix unitarization schemes. In Sec. 3 the corresponding multiplicity distributions of secondaries produced in the central rapidity region are calculated assuming that the distribution of particles produced by one cut Pomeron obeys the Poisson law. The results are compared with that observed at the LHC at the largest energy $\sqrt s=13$ TeV. We conclude in Sec. 4.

\section{AGK rules}

According to the AGK rules \cite{AGK} the contribution of the diagram with $n$ Pomerons to the multiparticle production cross section is obtained by cutting $k$ Pomerons. Each cut Pomeron produces some number of secondaries uniformly distributed in rapidity space. That is, writing the forward amplitude as
\begin{equation}
\textnormal{Im}{\cal A}_{(\textnormal{cut}\,\Bbb P)}(s,t=0)=s\sum_n C_n(-1)^{n-1} {\Bbb P}^n  
\end{equation}  
(${\Bbb P}$ here represents the Re$\Omega={\Bbb P}$ or Im$\chi={\Bbb P}/2$ in (\ref{eik},\ref{u}))\footnote{In the cut Pomeron we deal with discontinuity ($disc=2\mbox{Im}A$) while the uncut Pomeron can be to the left or the right of the cut (this is the origin of factor 2 in (\ref{agk})) and this way the real part canceled.} and cutting $k$ Pomerons from the term $C_n(-1)^{n-1}{\Bbb P}^n$ we get the cross-section 
$\sigma_n^k=2C_nc^k_n{\Bbb P}^n$,
where
\begin{equation}
\label{agk}
c^{k\neq 0}_n = (-1)^{n-k}2^{n-1}\frac{n!}{k!(n-k)!}\ ,
\end{equation}
\begin{equation}
c^{k=0}_n=(-1)^n(2^{n-1}-1)\ .
\end{equation}
The AGK factor $c^{k=0}_n$ corresponds to the cut between the Pomerons (as that in Fig.\ref{f1}a) when no secondaries are produced in the central region.\footnote{Using the identity $\sum_{k=0}^n(-1)^kn!/(k! (n-k)!)=0$ it is easy to check that the sum over all possible cuts $\frac 12\sum_k\sigma^k_n=C_n(-1)^{n-1}{\Bbb P}^n$ is equal to the $n$-Pomerons exchange contribution to the total cross-section. Strictly speaking, the first term, $2^{n-1}$ in $c^{k=0}_n$ should be multiplied by ${\Bbb P}$ (i.e. by the imaginary part of the one Pomeron exchange amplitude) while in the last term, $1$, corresponding to the cut going to the left (or right) of all the Pomerons we must keep the whole complex amplitude.}

\begin{figure*}[t]
\begin{center}
\vspace{-7.9cm}
\hspace{0.9cm}
\includegraphics[scale=0.75]{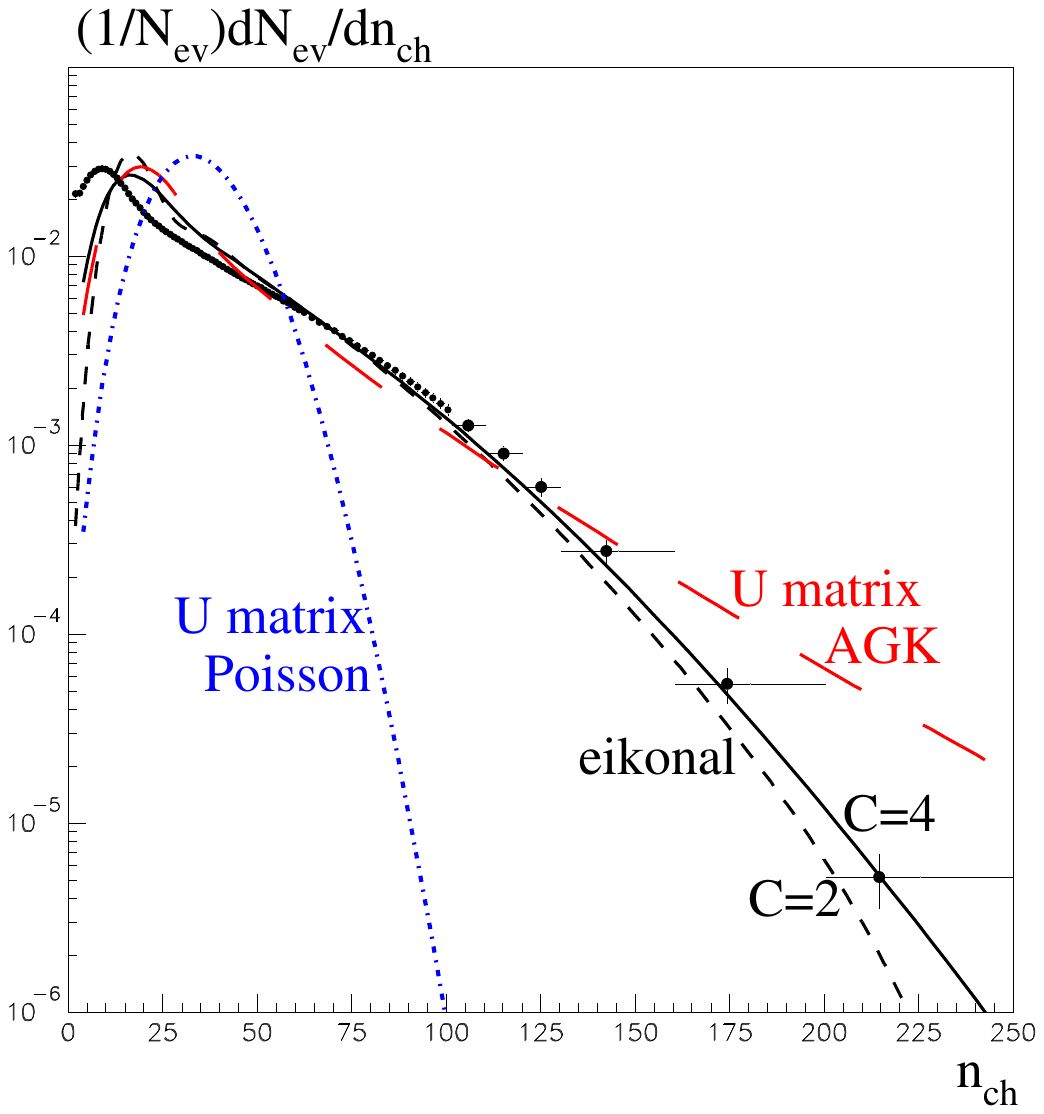}
\vspace{-0.5cm}
\caption{\sf Charged particle central region ($-2.5<\eta<2.5$) multiplicity distribution in the eikonal  (black continuous and short dashed curves) and the $U$-matrix (red long dashed and blue dot-dashed curves; $C=4$) unitarization schemes. The data are from~\cite{atl-epg}.
}
\label{f3}
\end{center}
\end{figure*}

That is for the process with the multiplicity $k$ times larger than that for the individual Pomeron we get the cross-section  
\begin{equation}
\label{k-cut}
\sigma^k (s,b) = 2\sum_nC_n\cdot (-1)^{n-k}2^{n-1}\frac{n![{\Bbb P}(s,b)]^n}{k!(n-k)!} .
\end{equation}

Note that, since the expansion in powers of ${\Bbb P}$ in (\ref{eik},\ref{u}) is of an alternating sign, the $k$ cut Pomerons contribution coming from the $n$ Pomeron exchange can be negative (see the factor $(-1)^n$ in (\ref{agk})). In such a case it describes the screening correction to the process with $k$ cut Pomerons coming from the term of a smaller $n$.\\ 

The factor $n!{\Bbb P}^n/(n-k)!$ in (\ref{k-cut}) can be written as the
derivative over ${\Bbb P}$ taken $k$ times.
\begin{equation}
\label{der}
\frac{n!{\Bbb P}^n}{(n-k)!} = {\Bbb P}^k\left(\frac{d}{d{\Bbb P}}\right)^k {\Bbb P}^n\ .
\end{equation}

Thus, putting $C_{n}=C_{n}^{eik}$ in (\ref{k-cut}), the cross sections $\sigma^k(s)=\int d^{2}b\, \sigma^k(s,b)$ for $k\geq 1$ read
\begin{equation}
\label{eik-s}
\sigma^k_{eik}(s) = \int d^2b\, \frac{[\mbox{Re}\Omega(s,b)]^k}{k!} 
\exp(-\mbox{Re}\Omega(s,b)) 
\end{equation}
in the eikonal case.

In a similar way, putting $C_{n}=C_{n}^{U}$ in (\ref{k-cut}), the $U$-matrix cross sections $\sigma^k_U(s)$ can be conveniently expressed as
\begin{equation}
\label{u-s}
\sigma^k_U(s) = 2\int d^2b\, \left[\frac{\mbox{Im}\hat{\chi}(s,b)}
{1+\mbox{Im}\hat{\chi}(s,b)}\right]^k\frac 1{1+\mbox{Im}\hat{\chi}(s,b)}\ .
\end{equation}

The problem however is that the expression (\ref{u-s}) contradicts the original amplitude (\ref{u}). From the  unitarity equation (\ref{un}) and (\ref{u}) we have 
\begin{eqnarray}
\label{uamp}
G_{inel}(s,b) &=& 2\mbox{Im}{\cal A}(s,b)-|{\cal A}(s,b)|^2 \nonumber \\
 &=& \frac{2\mbox{Im}\hat{\chi}(s,b)}{(1-i\hat{\chi}(s,b)/2)(1+i\hat{\chi}^{*}(s,b)/2)},
\end{eqnarray} 
while from (\ref{u-s}) we get
\begin{equation}
\label{uagk}
G_{inel}(s,b) = \sum_k\sigma^k_U(s,b)=2\frac{\mbox{Im}\hat{\chi}(s,b)}
{1+\mbox{Im}\hat{\chi}(s,b))}\ .
\end{equation}
In particular, at very large $\hat{\chi}\to\infty$ according to  (\ref{uamp}) $G_{inel}\to 0$ whereas from (\ref{uagk}) $G_{inel}\to 2$.

This indicates that the $U$-matrix unitarization is inconsistent with the AGK rules.\\

Indeed, the $U$-matrix unitarization corresponds to the ``quasi potential'' situation when the few short time (instant) interactions happen one after another, like as it was considered in the AFS paper~\cite{afs} and is shown in Fig.\ref{f2}b. On the other hand, it was demonstrated by Mandelstam~\cite{Mand} that at very high energies such a contribution vanishes and only the diagrams where all the Pomeron exchanges take place simultaneously (as e.g. in Fig.\ref{f1} or Fig.\ref{f2}a)
 survive. In other words, in the $U$-matrix case we cannot parametrize the function $\hat{\chi}(s,b)$ by the Pomeron exchange. One has to propose some other dynamics.  \\
 
Coming back to Fig.\ref{f2}b, we can say that in this kinematics one can cut only {\em one} object; it is impossible to cut a few "quasi potentials" (Pomerons) simultaneously. This way we will reproduce the result (\ref{uamp}) expected from (\ref{un},\ref{u}). However
 then there will be {\rm no} long-range (in rapidity) correlations and the obtained multiplicity distribution should be Poisson-like.

\section{Numerical estimates}
For numerical estimates, we take the parameters of the Pomeron trajectory and the Pomeron-proton coupling, $\beta(t)$,  from \cite{LMP} for the simplest version of fit without the Odderon and with the exponential form of $\beta(t)\propto \exp(Bt)$ (see Table I of \cite{LMP}). In both the eikonal and the $U$-matrix schemes, this provides a good description of the ATLAS LHC data: 
$\chi^2/DoF=0.86$ for the eikonal and $\chi^2/DoF=0.85$ for the $U$-matrix (using the parameters obtained in \cite{LMP} fitting to TOTEM data we get practically the same multiplicity distributions).

For the distribution generated in the central region by a single  Pomeron, we take the Poisson law. However here we have to account for the possibility of short-range (in rapidity) correlations. 
First, this is the electric charge conservation. In other words, actually, the new secondaries are created in pairs, say $\pi^+$ and $\pi^-$ separated by a relatively small rapidity interval.
 That is we have to expect the Poisson distribution over some number $N=N^{{\Bbb P}}_{ch}/2$ ($N^{{\Bbb P}}_{ch}$ is the number of charged particles generated by one Pomeron). 
 Besides this, the particles can be produced via the resonance or the mini-jet decay. Denoting the mean charged multiplicity of such a cluster by $C$ we have to expect the Poisson in a number of clusters $N=N^{{\Bbb P}}_{ch}/C$. Due to electric charge conservation, we expect $C>2$.\\
 
 The results corresponding to eikonal unitarization (\ref{eik-s}) are presented in Fig.\ref{f3} by the black lines for $C=2$ and $C=4$. The value of $N^{{\Bbb P}}_{ch}$ is chosen to reproduce the particle density $dN_{ch}/d\eta$ observed in the same paper~\cite{atl-epg}. The dependence on the $C$ value is not strong. Actually, the form of the curve is controlled by the distribution over the number of cut Pomerons, $k$. Using the $U$-matrix $k$ distribution (\ref{u-s}) we get the red (long dashed) curve which falls with $n_{ch}$ too slowly ($n_{ch}$ is the charged particle multiplicity observed in the central, $|\eta|<2.5$, region).
 
However, as was discussed at the end of Sec. 2 in the $U$-matrix case we have a possibility to cut only one Pomeron.
That is here we have to expect just the Poisson distribution shown in Fig.\ref{f3} by the blue (dot-dashed) line.\footnote{For both $U$-matrix curves we put $C=4$.} 
 
Note that since our simplified model does not account for the high mass diffractive dissociation (described by the triple-Pomeron and more complicated enhanced multi-Pomeron diagrams) it is not surprising that we do not reproduce the region of low multiplicities, $n_{ch}<20$.

So far, no one has successfully summed up {\em all} multi-Pomeron diagrams. It is expected that these interactions will mainly reduce the effective Pomeron intercept.

On the other hand, in our calculations, we had used not the theoretical intercept of the original Pomeron but the value coming from fitting $\sigma_{tot}$ and $d\sigma_{el}/dt$ cross sections within the eikonal (or $U$-matrix) approach. That is, this effect has already been accounted for. It is important to note that in any case, the eikonal or the $U$-matrix is the last step of any unitarization where some more complicated two-particle irreducible elastic amplitude plays the role of the Pomeron. 

The most acceptable and self-consistent model that takes into account Pomeron-Pomeron interactions was developed by S. Ostapchenko (see \cite{ostapchenko1} and \cite{ostapchenko2} for the last QGSJETIII version). This model, based on the eikonal approach, satisfactorily reproduces the data. However, it does not account for multiplicity distributions.

As far as we know, there is no generalization of the $U$-matrix approach that includes Pomeron-Pomeron interactions.

The most promising option at this stage is to explore the multi-channel eikonal (and $U$-matrix) using the Good-Walker formalism \cite{GW}. We plan to conduct this analysis in our next paper.

\section{Conclusion}
 
Using the AGK cutting rules~\cite{AGK} we calculated the multiplicity distribution of charged particles created by the multi-Pomeron diagrams at the highest LHC energy in the central rapidity region, $|\eta|<2.5$. The multi-Pomeron diagrams are considered within the eikonal or the $U$-matrix unitarization schemes.
 The results are compared with the ATLAS 13 TeV data~\cite{atl-epg}. As it is seen in Fig.\ref{f3} the eikonal approach better agrees with the data.
 
 Moreover, it is shown that the $U$-matrix scheme cannot be used for the unitarization of the Pomeron with $\alpha_{\Bbb P}(0)>1$ since it is inconsistent with the original elastic amplitude (\ref{u}). From the AGK rules at asymptotically high energy we get $G_{inel}(s,b)\to 2$ (\ref{uagk}) while from the (\ref{u}) and (\ref{un}) $G_{inel}(s,b)\to 0$. Another argument against the $U$-matrix approach was given in~\cite{KMR}, where it was shown that the $t$ channel unitarity generates the inelastic contribution ($G_{inel}$) which does not decrease at $s\to\infty$.

\section*{Acknowledgments}
 
The authors thank V.A. Khoze for reading the manuscript and the discussion. This research was partially supported by the Conselho Nacional de Desenvolvimento Cient\'{\i}fico e Tecnol\'ogico under Grant No. 307189/2021-0.

\end{document}